\documentclass{elsart}
\usepackage{natbib}
\usepackage{graphicx}
\usepackage{amssymb}

\begin{document}

\begin{frontmatter}

\title{Recent Achievements on the Development of the HERSCHEL/PACS Bolometer
arrays}

\author[label1]{Billot N.\corauthref{cor1}}
\ead{nbillot@cea.fr}
\corauth[cor1]{Corresponding author: Tel.: +33 (0)169 089 570; Fax: +33 (0)169 086 577}
\author[label2]{Agn\`ese P.}
\author[label1]{Boulade O.}
\author[label2]{Cigna C.}
\author[label1]{Doumayrou E.}
\author[label1]{Horeau B.}
\author[label1]{Lepennec J.}
\author[label1]{Martignac J.}
\author[label2]{Pornin J.-L.}
\author[label1]{Reveret V.}
\author[label1]{Rodriguez L.}
\author[label1]{Sauvage M.}
\author[label2]{Simoens F.}
\author[label1,label3]{Vigroux L.}
\address[label1]{CEA/Saclay/SAp, UMR CEA/CNRS/UP7 Laboratoire AIM, 
Bat. 709, l'Orme des merisiers, 91191 Gif-sur-Yvette, France}
\address[label2]{CEA/LETI Grenoble, 17 Avenue des Martyrs, 38054 Grenoble,
France}
\address[label3]{Institut d'Astrophysique de Paris, 75014 Paris, France}

\begin{abstract}
A new type of bolometer arrays sensitive in the far Infrared and
Submillimeter range has been developed and manufactured by
CEA/LETI/SLIR since 1997. These arrays will be integrated in the PACS
instrument (Photodetector Array Camera and Spectrometer) of ESA's
Herschel Space Observatory (launch date 2007). The main innovations of
CEA bolometers are their collective manufacturing technique
(production of 3-side buttable 16x16 arrays) and their high
mapping efficiency (large format detector and instantaneous Nyquist
sampling). The measured NEP is $2.10^{-16}\,W/\sqrt{Hz}$ and the
thermometric passband about $4-5\,Hz$. In this article we describe
CEA bolometers and present the results obtained during the last
test campaign.
\end{abstract}


\end{frontmatter}

\section{Introduction}
\label{S0}

The Herschel Mission stands as the next great step in FIR and
submillimeter astronomy. This huge satellite will hold three
scientific instruments (HIFI, SPIRE and PACS \cite{1})
and the largest telescope ever sent in space (3.5m in diameter). It
will explore a universe that is unobservable from the ground. The
space observatory will carry out large scale surveys of the distant
Universe to study galaxy evolution. It will also focus on our own
Galaxy to inform us on the composition, chemistry and life-cycle of
the interstellar medium. Targets like comets, solar system planets and
extra-solar planetary disks are also foreseen.  The Herschel satellite
is due to launch in 2007 to reach the second Lagrangian point L2 after
a 6 months journey. The mission is funded by the European Space Agency
and has a nominal duration of 3.5 years. CEA/DAPNIA and CEA/LETI have
developed a very sensitive camera made up of more than 2500
micro-bolometers (operating at 300 mK) for use in the PACS photometer
(60-210 microns).\newline During summer 2005, we have run performance
measurements at CEA Saclay to qualify the Bolometer Focal Plane (BFP)
before delivering it to Max-Planck-Institut fur extraterrestrische Physik (MPE) 
Garching by the end of the year for
integration in the PACS instrument.\newline Section \ref{S1} is an
outline of the detector design while section \ref{S2} concentrates on
the latest results obtained during the qualifying test campaign.

\section{PACS bolometers}
\label{S1}
\begin{figure}[t]
\begin{center}
\includegraphics[width=12cm]{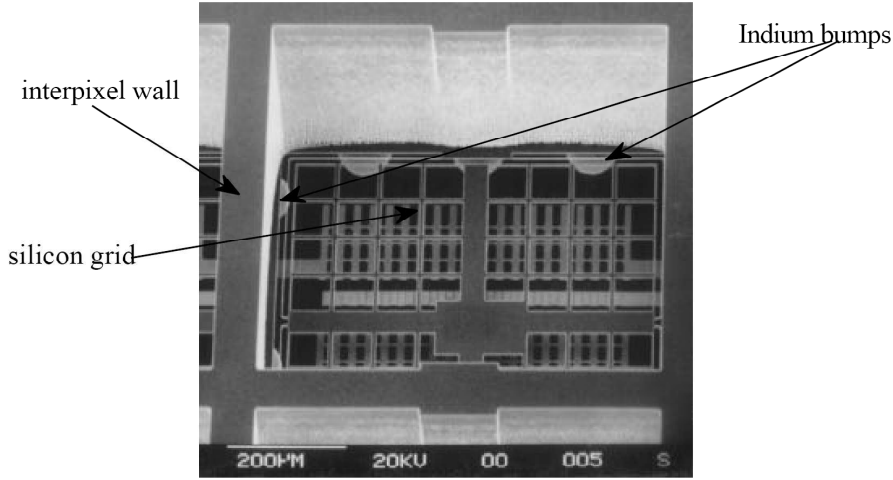}
\caption{Microphotography of a single pixel ($750\times750 \mu m$). }
\label{fig:pixel}
\end{center}
\end{figure}

PACS photometer uses a dichroic to split light between two channels
dedicated to short- ($60-130 \mu m$) and long-wavelengths ($130-210
\mu m$). The two focal planes have the same field of view ($3.5\times
1.75$ arcmin) and consist of tiled matrices manufactured by blocks of
16x16 pixels and for the first time in bolometers' history, the cold
readout electronics circuit (300 mK) includes a 16 to 1 multiplexing
function. The detector is based on an all-Silicon technology
development that takes advantage of Si micro-machining techniques
maturity.\newline Unlike most current submillimetric imagers, CEA
bolometers do not use Winston cones as light concentrators but rather
a reflecting plate and a quarter-wave cavity to optimise
absorption. All these features made it possible to build detectors with
a large number of small contiguous pixels. Actually twelve of these
pixels fit in the Airy disc of the telescope (pixel width = $750\mu m$
= $0.5F\lambda$); which is, according to sampling theory, enough to
correctly sample the Point Spread Function (PSF) in a single shot (no need for jiggling
observing mode). This makes the camera very efficient at mapping large
areas of the sky.\newline The simplest way to describe the PACS
bolometers is to divide each pixel in a detection layer which
comprises the suspended absorbing grid, the thermometric sensor and
the Si interpixel wall on the one hand ; and a readout layer that
contains the CMOS multiplexer and the reflecting gold sheet on the
other hand. The two layers are hybridised with $20\mu m$ Indium bumps
that ensure electrical and thermal contacts (see figure
\ref{fig:pixel}). The bolometer signal is read at the middle point of
a resistor bridge; one resistor is located in the interpixel wall and
held at bath temperature ($\sim$ 300mK) and the other is implanted in
a mesa configuration in the middle of the absorber.
\begin{description}
\item[Thermometers] are very high impedance resistors (about
$4\rm{G}\Omega$) working in the `hopping conduction' regime at
300mK. Such a high impedance is achieved by ion implantation (100\%
phosphorus compensated with 50\% boron) on a double
Silicon-On-Insulator (SOI) substrate. They exhibit an exponential
dependence on temperature. Non-ohmic effects are minimised for long
linear thermometer geometry.
\item[EM absorption] takes places at the top of a quarter-wave
resonant cavity and reaches nearly 100\% efficiency when the absorber
(TiN deposited on Si grid) is put where the antinode of the
stationary wave is created. The size of the cavity is tuned with the
Indium bumps diameter ($20 \mu m$ in this case).
\item[The cold electronics] is located just below the $\lambda/4$
cavity reflector to minimise stray capacitances from electronics
lines. A CMOS multiplexer (inspired by ISOCAM technology \cite{2}) is also
implemented to reduce dissipation at the 300mK stage and increase the
potential number of pixels in the confined focal plane. It offers a 16
to 1 MUX function.
\end{description}
More details on the detector design or on mechanical and thermal
issues may be found in \cite{3}.\newline Herschel's telescope
will be passively cooled to 80K and should have a relatively low
emissivity (about 4\%). The blackbody emission from the dish will
represent most of the flux illuminating the detector and we expect
astronomical sources to be a percent or less of the background
flux. The requirement on sensitivity (photon noise limited) is 5mJy
(5$\sigma$, 1 hour) which is equivalent to a Noise Equivalent Power
(NEP) of $ \sim 1.5\times 10^{-16} \;W/\sqrt{Hz}$, depending on
efficiency assumptions.

\section{Performance tests}
\label{S2}
\begin{figure}[htb]
\begin{center}
\includegraphics[width=12cm]{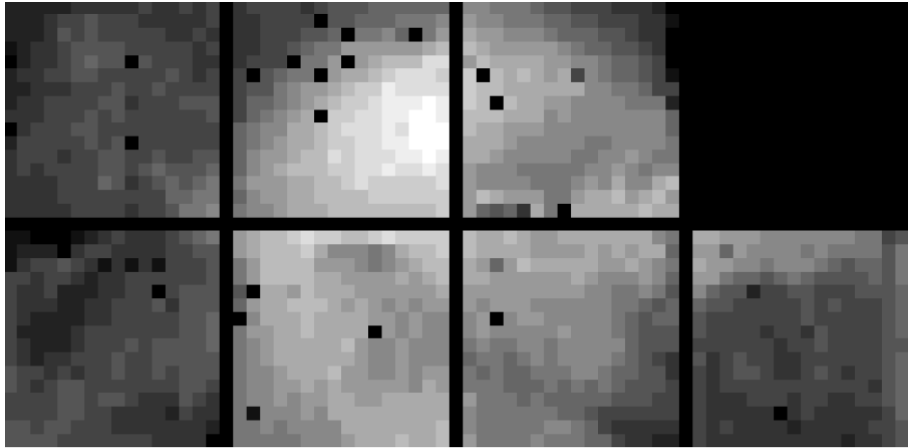}
\caption{Responsivity map of the short-wavelength BFP (2pW/pixel
background and 500fW modulation). The circular central feature in the
map is due to internal reflexions inside the cryostat. Finer
measurements will be done at MPE using the two Internal Calibration
Sources inside PACS. Dead pixels represent less than 2\% of the total
number of pixels. The missing array was not functional at the time of
the test but has been repaired and is currently being tested.}
\label{fig:map_resp}
\end{center}
\end{figure}

 The experimental setup in Saclay consists in a cryostat, an $^3$He
sorption cooler and two regulated blackbodies. The illumination of the
detector is usually modulated between the two blackbodies to mimic
chopped observations. The background flux is set to 2pW/pixel
corresponding to a telescope at 80K.\newline During the last test
campaign we quantified the noise level, the responsivity and the
thermometric passband. We carried out systematic measurements of these
parameters as a function of the bias voltage applied across the
resistor bridge for two different background fluxes. All measurements
were obtained in the nominal readout mode (differential readout). The
signal was sampled at the nominal readout frequency of 40Hz.
\begin{description}
\item[The noise spectral
density] exhibits two distinct components: a low-frequency noise
spectrum due to electronics and temperature drifts and a white noise
spectrum at higher frequencies. The $1/f$ knee frequency occurs at
about 0.1-1 Hz and the noise level measured at 3Hz is about $9\mu
V/\sqrt{Hz}$. 
\item[The responsivity] is measured to be about $4\times10^{10}\;V/W$
with a background flux of 2pW/pixel and a modulation of 0.5 pW (figure
\ref{fig:map_resp}). Such high responsivities are required to overcome
the relatively high noise level due to the cold CMOS held at
300mK.
\item[The NEP] is actually a figure of
sensitivity and is defined as $\frac{noise}{responsivity}$. We find an
optimal NEP of $2.2\times10^{-16}\;W/\sqrt{Hz}$ for a voltage bias of
2.8V. 
\item[The thermometric passband] is defined as being the
modulating frequency at which the signal is attenuated by 3dB. It is a
crucial parameter that will constrain the maximum scanning speed of
the telescope. The measurements gave a passband of 4-5 Hz.
\end{description}
Since Herschel will be under the influence of ionising particles from
the Solar wind, the detector was also tested for irradiation. It
received a cumulated dose of 20krad of $\gamma$-rays ($\sim$10 years
in space) and different fluxes of protons (20MeV) and alpha particles
(30MeV) without degradation to the performances (upset time $< 200$ms).

\section{Prospects}
\label{S3}
The next step is the PACS calibration campaign that will take place
at MPE Garching for 6 months before delivery to ESA in mid-2006.
Meanwhile, CEA bolometers are being optimised for longer wavelengths
to match atmospheric windows. This way, they could be used on ground
telescopes like KOSMA, APEX or IRAIT at Concordia Station in
Antarctica \citep{4}.

\end{document}